\documentstyle[amsfonts,12pt,oldlfont]{article}

\title{The solution to the q-KdV equation}

\author{M. Adler\thanks{Department of Mathematics,
Brandeis University, Waltham, Mass 02254,
USA. The support of a National Science Foundation grant \#
DMS-9503246 is gratefully acknowledged.}~~~~~E.
Horozov\thanks{Department of Mathematics and Informatics,
Sofia University, 5 J. Bourchier Blvd., Sofia 1126,
Bulgaria.}~~~~~P. van Moerbeke\thanks{Universit\'e de Louvain,
1348 Louvain-la-Neuve, Belgium and Department of Mathematics,
Brandeis University, Waltham, Mass 02254, USA. The  support of a
National Science Foundation grant \# DMS-9503246, a Nato, a FNRS
and a Francqui Foundation grant is gratefully acknowledged.
part of this work was carried out at the Centre Emile Borel,
Paris during fall 96.}}
\date{October 15, 1997}

\newcommand{\MAT}[1]{\left(\begin{array}{*#1c}}
\newcommand{\mat}{\end{array}\right)}

\newcommand{\qed}
{%
\mbox{}%
\nolinebreak%
\hfill%
\rule{2mm}{2mm}%
\medbreak%
\par%
}

\newcommand{\rg}{\rightarrow}

\newcommand{\DF}{\Longleftrightarrow}
\newcommand{\lrg}{\longrightarrow}

\newcommand{\BZ}{{\Bbb Z}}
\newcommand{\BC}{{\Bbb C}}

\newcommand{\iy}{\infty}
\newcommand{\pl}{\partial}
\newcommand{\al}{\alpha}

\newcommand{\om}{\omega}

\newcommand{\dt}{\delta}

\newcommand{\vr}{\varepsilon}

\newcommand{\BR}{{\Bbb R}}
\newcommand{\lb}{\lambda}
\newcommand{\Lb}{\Lambda}
\newcommand{\BX}{{\Bbb X}}

\def\span{\mathop{\rm span}}

\def\be{\begin{equation}}
\def\ee{\end{equation}}
\def\bea{\begin{eqnarray}}
\def\eea{\end{eqnarray}}

\ifx\undefined\Bbb
        \let\Bbb\bf
\fi
\ifx\undefined\curvearrowright
        
\else
        
\fi

\catcode`\@=11
\catcode`@=12
\begin{document}
\maketitle

\setcounter{equation}{0}

\noindent {\bf Abstract}: Let KdV stand for the $N$th
Gelfand-Dickey reduction of the KP hierarchy. The purpose of
this paper is to show that any KdV solution leads effectively to
a solution of the
$q$-approximation of KdV. Two different $q$-KdV approximations were
proposed, first
one by E. Frenkel \cite{F} and a variation by Khesin, Lyubashenko
and Roger \cite{KLR}. We show there is a dictionary
between the solutions of $q$-KP and the 1-Toda lattice
equations, obeying some special requirement; this is based on an
algebra isomorphism between difference operators and
$D$-operators, where
$Df(x)=f(qx)$. Therefore every notion about the 1-Toda lattice
can be transcribed into $q$-language.

\vspace{1cm}

Consider the $q$-difference operators $D$ and $D_q$, defined by
$$
Df(y)=f(qy)~~~\mbox{ and }
																					~~~D_qf(y):=\frac{f(qy)-f(y)}{(q-1)y},
$$
and the $q$-pseudo-differential operators
$$
Q=D+u_0(x)D^0+u_{-1}D^{-1}+...\mbox{
and }~~Q_q=D_q+v_0(x)D_q^0+v_{-1}(x)D_q^{-1}+...
$$
The following $q$-versions of KP were proposed by E. Frenkel
\cite{F} and a variation by Khesin, Lyubashenko and Roger
\cite{KLR}, for $n=1,2,...$:
\be
{\pl Q\over\pl t_n}=\bigl[\left(Q^n\right)_+,Q\bigr]
~~~~~~~~~~~~~~~ ~~~~~~~\mbox{{\em (Frenkel system)}}
\ee
\be
{\pl Q_q\over\pl
t_n}=\bigl[\left(Q_q^n\right)_+,Q_q\bigr],
~~~~~~~~~~~~~~~~~~~ \mbox{{\em (KLR system)}}
\ee
where $(~)_+$ and $(~)_-$ refer to the $q$-differential and
strictly $q$-pseudo-differential part of $(~)$. The two systems
are closely related, as will become clear from the isomorphism
between $q$-operators and difference operators, explained below. The
purpose of this paper is to give a large class of solutions to both
systems.

The $\delta$-function $\dt(z):=\sum_{i \in \BZ}z^i$;
enjoys the property
$f(\lb,\mu)\dt(\lb/\mu)=f(\lb,\lb)\dt(\lb/\mu)$. Consider an
appropriate space of functions
$f(y)$ representable by ``Fourier" series in the
basis $\varphi_n(y):=\dt(q^{-n}x^{-1}y)$ for fixed $q\neq 1$,
$$
f(y)=\sum_{-\iy}^{\iy}f_n\varphi_n(y);
$$
the operators $D$, defined by $Df(y)=f(qy)$, and multiplication by
a function $a(y)$ act on the basis elements, as follows:
$$
D\varphi_n(y)=\varphi_{n-1}(y)\mbox{ and }a(y)
\varphi_n(y)=a(xq^n)\varphi_n(y).
$$
Therefore, the Fourier transform,
$$f\longmapsto \hat f=(...,f_n,...)_{n\in \BZ},
$$
induces an algebra isomorphism, mapping $D$-operators onto
a special class of $\Lb$-operators in the shift
$\Lb:=\Bigl(\dt_{i,j-1}\Bigr)_{i,j\in\BZ}$, as follows:
\be
\sum_i a_i(y)D^i\longmapsto
\sum_i\mbox{ diag}\left(
...,a_i(xq^n),...\right)_{n
\in \BZ}\Lb^i;
\ee
conversely, any difference operator, depending on
$x$, of the type (0.3) i.e., annihilated
by $D-Ad_{ \Lb}$, where $(Ad_{\Lb})a=\Lb a \Lb^{-1}$, maps into
a $D$-operator. This is the crucial basic isomorphism used
throughout this paper.

To the shift
$\Lb$ and to a fixed diagonal matrix
$\lb=
$ diag$(...,\lb_{-1},\lb_0,\lb_1,...)$, we associate new operators
$$\tilde\Lb=-\lb\Lb\quad\mbox{and}\quad\widetilde{\tilde\Lb}
=\tilde\Lb+\lb=-\lb(\Lb-1).
$$
Observe that, under the isomorphism (0.3),
$$D\longmapsto\Lb~, ~~
\frac{1}{(q-1)x}D\longmapsto\tilde\Lb\quad\mbox{and}\quad
D_q \longmapsto \widetilde{\tilde\Lb},
$$
upon setting $\lambda^{-1}_n = (1-q) x q^{n-1}$.

Defining the simple vertex operators
\be
X(t,z):=e^{\sum_1^{\iy}t_iz^i}e^{-\sum_1^{\iy}
\frac{z^{-i}}{i}\frac{\pl}{\pl t_i}},
\ee
we now make a statement concerning the so-called full one-Toda
lattice; the latter describes deformations of of a bi-infinite
matrix $L$, which is lower-triangular, except for a diagonal
and a constant subdiagonal just above the main diagonal. The
{\em first formula} (0.6) below gives a solution to the
{\em Frenkel system} (Theorem 0.1), upon replacing $\tilde \Lb$
by
$\Lb$, whereas the {\em second formula} (0.6) gives a solution to
the {\em KLR system} (Theorem 0.2). Note, Theorem 0.1 is given
for arbitrary
$\lb=(...,\lb_{-1},\lb_0,\lb_1,...)$.
We shall need the well-known Hirota symbol for a polynomial $p$,
$$
p(\pm\tilde\pl)f\circ
g:=p\left(\pm\frac{\pl}{\pl y_1},\pm\frac{1}{2}\frac{\pl}{\pl
y_2},...\right)f(t+y) g(t-y)\Biggl|_{y=0}.
$$
Note $A_+$ refers to the upper-triangular part of a matrix
$A$, including the diagonal, and for $\al \in \BC$, set
$[\al]:=(\al,\frac{1}{2}\al^2,\frac{1}{3}\al^3,...)\in
\BC^{\iy}$.

\vspace{0.5cm}

\proclaim Theorem 0.1. Given an integer $N\geq 2$, consider an
arbitrary $\tau$-function for the KP equation such that
$\pl\tau/\pl t_{iN}=0$ for $i=1,2,3,...$ ($N$-KdV hierarchy). For a
fixed
$\lb, ~\nu,~c\in\BC^{\iy},$ the infinite sequence of
$\tau$-functions \footnote{$\tau_n$ for $n<0$ is
defined later in (3.3).}
$$
\tau_{ n}: =e^{\nu_{ n}\sum_{i=1}^{\iy} t_{iN}} X
(t,\lambda_{n}) ...X (t,\lambda_{1})
\tau (c+t),\quad\tau_0=\tau(c+t),~ \mbox{ for } n\geq 0;
$$
satisfies the 1-Toda bilinear identity for all $t,t'\in
\BC^{\iy}$ and all $n>m$:
$$
\oint_{z=\iy}\tau_n(t-[z^{-1}])\tau_{m+1}(t'+[z^{-1}])
e^{\sum_1^{\iy}(t_i-t'_i)z^i}
z^{n-m-1}dz=0.
$$
The bi-infinite matrix (a full matrix below the main diagonal),
where $p_{\ell}$ are the elementary Schur polynomials,
\be
L=\sum_{\ell=0}^{\iy}\frac{p_{\ell}(\tilde\pl)
\tau_{n+2-\ell}\circ\tau_n}
{\tau_{n+2-\ell} \tau_n}\tilde\Lb^{1-\ell}
\ee
has the following properties:
\medbreak\indent {\rm (i)} $L^N$ satisfies the $1$-Toda
lattice
$$
{\pl L^N\over\pl t_n}=\bigl[\left(L^n\right)_+,L^N\bigr],\quad
n=1,2,..,
$$
\medbreak\indent {\rm (ii)} $L^N$ is \underline{upper triangular}
and admits the following expression in
terms\footnote{in the expressions below, the coefficients of
the $\widetilde{\tilde\Lb}$'s are diagonal matrices, whose $0$th
component is given by the expression appearing below; i.e.,
$\sum_1^{N} b_j$ stands for diag$(\sum_1^{N} b_{j+n})_{n \in
\BZ}$}
of $\tilde\Lb$ and $\widetilde{\tilde\Lb}$:
\bea
L^N&=&\tilde\Lambda^N+\sum_1^N(\lb_j+b_j)\tilde\Lb^{p-1}+
\Biggl(\sum_0^{N-1}a_j+
\sum_{1\leq i\leq j\leq
N-1}(\lb_i+b_i)(\lb_j+b_j)\Biggr)\tilde\Lb^{N-2}\nonumber\\
& &\hspace{8cm}+...+(\nu_{n+1}
-\nu_n)_{n\in\BZ}\tilde\Lb^0\nonumber\\
&=&\widetilde{\tilde\Lb}^N+\left(\sum_1^{N}b_j\right)
\widetilde{\tilde\Lb}^{N-1} \nonumber\\
& &\hspace{1.5cm}+\left(\sum_0^{N-1}a_j-\sum_1^{N-1}
(b_{N}-b_i)
\lb_i+\sum_{1\leq i\leq j\leq
N-1}b_ib_j\right)\widetilde{\tilde\Lb}^{N-2}+...
\eea
with
\be
b_k=\frac{\pl}{\pl
t_1}\log\frac{\tau(c+t-\displaystyle{\sum_1^{k}}[\lb_i
^{-1}])}{\tau(c+t-\displaystyle{\sum_1^{k-1}}[\lb_i^{-1}])},~~~~
a_k=\left(\frac{\pl}{\pl
t_1}\right)^2\log\tau\left(c+t-\sum_1^k[\lb_i^{-1}]\right),
\ee
for $k\geq 1$. These expressions for $k\leq 0$
will be given in (3.4) and (3.5).

In view of (0.6), the shift
$$\Lb:~b_k  \longmapsto \Lb b_k=b_{k+1}~~~\mbox{and}~~~ a_k
\longmapsto \Lb a_k=a_{k+1}
$$
corresponds to the following transformation,
\be
\Lb:~c\mapsto c-[\lb_1^{-1}]~~~\mbox{and}~~~\lb_i \mapsto
\lb_{i+1}.
\ee
Therefore, in order that $L^N$ satisfies the form of the right
hand side of (0.3), we must make
$c$ and
$\lb_i$ depend on
$x$ and
$q$, such that the map $\Lb$ on $a,~b,~ \lb$ corresponds to
$D$, in addition to the fact that all
$\lb_i$ must tend to $\iy$ simultaneously and $c$ to $(x,0,0,...)$,
when $q$ goes to $1$. So, $c$ and $\lb$ must satisfy:
\be
\left
\{\begin{array}{l}
Dc(x) =c(x) - [\lambda^{-1}_1]\\
D \lambda_n = \lambda_{n+1}\\
\lim_{q \to 1} \lb_i=\iy\\
\displaystyle{\lim_{q \to 1}} \,c (x)=\bar x:=(x,0,0,...);
\end{array}
\right.
\ee
its only solution is given by:
\be
\lambda^{-1}_n = (1-q) x q^{n-1} \ \ \ \mbox{and} \ \ \ c(x) =
\left({(1-q) x \over 1-q} , {(1-q)^2 x^2 \over 2(1-q^2)} ,
{(1-q)^3 x^3 \over 3 (1-q^3)}, ... \right),
\ee
and thus $D^n c(x)=c(x)-\sum_1^n[\lb_i^{-1}]$. With this choice
of $\lb_n$,
\be
\frac{1}{(q-1)x}D\longmapsto\tilde\Lb\quad\mbox{and}\quad
D_q:=\frac{D-1}{(q-1)x} \longmapsto \widetilde{\tilde\Lb}.
\ee
In analogy with (0.4), we define the simple $q$-vertex
operators:
\be
X_q(x,t,z):=e_q^{xz}X(t,z)~~\mbox{and}~~
\tilde X_q(x,t,z):=\left(e_q^{xz}\right)^{-1}X(-t,z).
\ee
in terms of (0.4) and
the $q$-exponential $e_q^x:= e^{\sum_1^{\iy}
\frac{(1-q)^k x^k}{k(1-q^k)} }$.
Therefore under the isomorphism
(0.3), Theorem 0.1 can be translated into $q$-language, to
read:

\proclaim Theorem 0.2. Any KdV $\tau$-function leads to a
$q$-KdV $\tau$-function $\tau(c(x)+t)$; the latter satisfies the
bilinear relations below, for all $x\in \BR,~t,t'\in \BC^{\iy}$,
and all $n> m$, which tends to the standard KP-bilinear identity,
when $q$ goes to $1$:
\begin{eqnarray}
& &\oint_{z=\iy}D^n
\left(X_q(x,t,z)\tau(c(x)+t)\right)D^{m+1}\left(\tilde
X_q(x,t',z)\tau(c(x)+t')\right) dz=0 \nonumber\\
& &~~~~~~\longrightarrow ~~\oint_{z=\iy}
X(t,z)\tau(\bar x+t)~
X(t',z)\tau(\bar x+t')~dz=0\nonumber\\
\end{eqnarray}
Moreover, the
$q$-differential operator $Q_q^N$ has the form below and tends to
the differential operator
${\cal L}^N$ of the KdV hierarchy, when $q$ goes to $1$:
\begin{eqnarray}
Q_q^N&=&D_q^N+\frac{\pl}{\pl t_1}
\log\frac{\tau(D^{N}c+t)}{\tau(c+t)}D_q^{N-1} \nonumber\\
& &\quad +\,\Biggl(\sum_{i=0}^{N-1}\frac{\pl^2}{\pl
t_1^2}\log\tau(D^i c+t) \nonumber \\
& & \quad -\sum^{N-2}_{i=0}\lb_{i+1}\left(\frac{\pl}{\pl
t_1}\log\frac{\tau(D^{N}c+t)}{\tau(D^{N-1}c+t)}-\frac{\pl}{\pl
t_1}\log
\frac{\tau(D^{i+1}c+t)}{\tau(D^i c+t)}\right) \nonumber \\
 & &\quad +\sum_{0\leq i \leq j \leq
N-2}\frac{\pl}{\pl
t_1}\log\frac{\tau(D^{i+1}c+t)}{\tau(D^i c+t)}\frac{\pl}{\pl
t_1}\log
\frac{\tau(D^{j+1}c+t)}{\tau(D^j
c+t)}\Biggr)D_q^{N-2}+...\nonumber \\
 && \nonumber \\
& &\longrightarrow \left(\frac{\pl}{\pl
x}\right)^N+N\frac{\pl^2}{\pl t_1^2}\log
\tau(\bar x +t)\left(\frac{\pl}{\pl
x}\right)^{N-2}+...~.
\end{eqnarray}

M.A. and PvM thank Edward Frenkel for kindly discussing this
problem during spring 1996. For a systematic study of discrete
systems, see Kupershmidt \cite{K} and Gieseker \cite{G}.

\section{The KP $\tau$-functions and Grassmannians}

\setcounter{equation}{0}

KP $\tau$-functions satisfy the differential Fay
identity for all $y,z \in \BC$, in terms of the Wronskian
$\{f,g\}:=f'g-fg'$, as shown in \cite{AvM1,vM}:
$$
\{\tau(t-[y^{-1}]),\tau(t-[z^{-1}])\}
\hspace{7cm}
$$
\be\hspace{3cm} +(y-z)(\tau(t-[y^{-1}])
\tau(t-[z^{-1}])-\tau(t)
\tau(t-[y^{-1}]-[z^{-1}])=0.
\label{9}
\ee
In fact this identity characterizes the $\tau$-function, as shown
in \cite{T}. We shall need the following, shown in \cite{AvM1}:

\proclaim Proposition 1.1. Consider $\tau$-functions $\tau_1$
and $\tau_2$, the corresponding wave functions
\be
\Psi_i = e^{\sum_{i\geq 1} t_i z^i} {\tau_i (t-[z^{-1}])\over
\tau_i (t)}=e^{\sum_{i\geq 1} t_i z^i}\left(1+O(z^{-1})\right)
\ee
and the associated infinite-dimensional planes, as points in the
Grassmannian $Gr$,
$$\tilde W_i=\mbox{span}\{\left(
\frac{\pl}{\pl t_1}\right)^k \Psi_i(t,z), \mbox{ for }
k=0,1,2,...\};$$ then the following statements are equivalent
\newline\noindent(i) $ z \tilde W_2 \subset \tilde W_1 $;
\vspace{0.2cm}\newline\noindent(ii) $ z \Psi_2 (t,z) =
\frac{\pl}{\pl t_1}
\Psi_1 (t,z) - \alpha \Psi_1 (t,z)$, for some function
$\alpha = \alpha (t)$;
\vspace{0.2cm}\newline\noindent(iii)\be\{ \tau_1
(t-[z^{-1}]),\tau_2 (t)\} + z (\tau_1 (t-[z^{-1}]) \tau_2 (t)
- \tau_2 (t-[z^{-1}])\tau_1 (t)) = 0 \ee
\newline\noindent When (i), (ii) or (iii) holds, $\alpha(t)$ is
given by
\be
\alpha (t) = {\pl \over \pl t_1} \log {\tau_2 \over \tau_1}.
\ee

Proof: To prove that (i) $\Rightarrow$ (ii), the inclusion
$z\tilde W_2 \subset \tilde W_1$ implies $z\tilde W^t_2 \subset
\tilde W^t_1$, where $\tilde
W^t=\tilde We^{-\sum_1^{\iy}t_iz^i}$; it follows that
$$
z \psi_2 (t,z) = z (1+O(z^{-1})) \in W_1^t
$$
must be a linear combination, involving the operator $\nabla =
{\pl \over \pl x} + z$ and the wave
functions $\Psi_i=e^{\sum_1^{\iy} t_i z^i}\psi_i$:
\be
z \psi_2 = \nabla \psi_1 -\alpha (t)  \psi_1,\mbox{ and thus }
z
\Psi_2 = {\pl \over \pl t_1} \Psi_1 - \alpha (t) \Psi_1.
\ee
The expression (1.4) for $\alpha (t)$ follows from equating the
$z^0$-coefficient in (1.5), upon using the $\tau$-function
representation (1.2). To show that (ii) $\Rightarrow$ (i), note
that
$$
z \Psi_2 = {\pl \over \pl t_1} \Psi_1 - \alpha \Psi_1 \in W^0_1
$$
and taking $z$-derivatives, we have
$$
z \left( \frac{\pl}{\pl t_1}\right)^j \Psi_2 =  \left(
\frac{\pl}{\pl t_1}\right)^{j+1} \Psi_1 + \beta_1 \left(
\frac{\pl}{\pl t_1}\right)^j \Psi_1 + \cdots + \beta_{j+1}
\Psi_1,
$$
for some $\beta_1,\cdots,\beta_{j+1}$ depending on $t$ only;
this implies the inclusion (i). The equivalence (ii)
$\Longleftrightarrow$ (iii) follows from a straight forward
computation using the $\tau$-function representation (1.2) of (ii)
and the expression for $\alpha (t)$.\qed

\section{The full one-Toda lattice}

\setcounter{equation}{0}

For details on this sketchy exposition, see
\cite{AvM5}. The one-Toda lattice equations
\be
\frac{\pl L}{\pl t_n}=[(L^n)_+,L],
\ee
are deformations of an infinite matrix
\begin{equation}
L=\sum_{-\iy<i\le 0}a_i\tilde\Lambda^i +\tilde\Lambda,\mbox{with}
~\tilde\Lambda:=\lb \Lb=\vr \Lb \vr^{-1},
\end{equation}
for diagonal matrices $\lb$ and $\vr$, with non-zero entries,
and diagonal matrices $a_i$, depending on $t=(t_1,t_2,\dots)$.
One introduces wave and and adjoint wave vectors $\Psi(t,z)$ and
$\Psi^{\ast}(t,z)$, satisfying
$$L\Psi=z\Psi~~\mbox{ and }~~
L^{\top}\Psi^*=z\Psi^*$$
and
\begin{equation}
\frac{\pl \Psi }{\pl t_n}=(L^n)_+
\Psi~~~~
\frac{\pl \Psi^*}{\pl t_n}=-((L^n)_+)^{\top} \Psi^*.
\label{1.3}
\end{equation}
The wave vectors $\Psi$ and $\Psi^*$ can
be  expressed in terms of one sequence of  $\tau$-functions
$\tau(n,t):=
\tau_n(t_1,t_2,\dots),\quad n\in\BZ$,
to wit:
$$
\Psi(t,z)=\left(e^{\sum^{\iy}_1 t_iz^i}
\psi(t,z) \right)_{n\in\BZ}=\left(
        \frac{\tau_n(t-[z^{-1}])}{\tau_n(t)}
e^{\sum^{\iy}_1 t_iz^i}\vr_n z^n
\right)_{n\in\BZ},
$$
\be
\Psi^*(t,z)=\left(e^{-\sum^{\iy}_1 t_iz^{i}}\psi^*(t,z)
\right)_{n\in\BZ}=\left(
\frac{\tau_{n+1}(t+[z^{-1}])}{\tau_{n+1}(t)}
e^{-\sum^{\iy}_1 t_iz^{i}}\vr_{n}^{-1}z^{-n}
\right)_{n\in\BZ}
\label{1.4}
\ee
It follows that, in terms of $\chi(z):=(z^n)_{n\in
\BZ}$:
$$\Psi=e^{\sum^{\iy}_1 t_iz^i} S\vr \chi(z),\quad\mbox{with}\quad
S=\frac{\sum_0^{\iy}p_n(-\tilde\pl)\tau(t)}{\tau(t)}\tilde\Lb^{-n},
$$
$$
\Psi^*=e^{-\sum^{\iy}_1
t_iz^i}(S^{\top})^{-1}\vr^{-1}\chi(z^{-1}), \quad\mbox{with}\quad
S^{-1}=\sum_0^{\iy}\tilde\Lambda^{-n}~\Lambda\frac{
 p_n(\tilde\pl)\tau(t)} {\tau(t)}
$$
and thus,
\bea
L&=&S\tilde\Lambda S^{-1} \nonumber \\
&=&\sum_{\ell=0}^{\iy}\frac{p_{\ell}(\tilde\pl)
\tau_{n+2-\ell}\circ\tau_n}
{\tau_{n+2-\ell} \tau_n}\tilde\Lb^{1-\ell}\nonumber \\
&=& \tilde\Lb+\left(\frac{\pl}{\pl
t_1}\log\frac{\tau_{n+1}}{\tau_n}\right)_{n\in\BZ}\tilde\Lb^0+\left(\left(
\frac{\pl}{\pl
t_1}\right)^2\log\tau_n\right)_{n\in\BZ}\tilde\Lb^{-1}+...
\end{eqnarray}

With each component of the wave vector $\Psi$, we associate a
sequence of infinite-dimensional planes in the Grassmannian $Gr$
\begin{eqnarray}
W_n&=&\mbox{ span}_{\BC} ~\{  \left( \frac{\pl}{\pl t_1}
\right)^k
\Psi_n(t,z),~~k=0,1,2,...\}\nonumber \\
&=&e^{\sum_1^{\iy}t_i z^i}\mbox{ span}_{\BC}~ \{  \left(
\frac{\pl}{\pl t_1}+z
\right)^k
\psi_n(t,z),~~k=0,1,2,...\}\label{1.6}
\end{eqnarray}
and planes
$$
W^\ast_n=\mbox{ span}_{\BC} ~\{  \left( \frac{\pl}{\pl t_1}
\right)^k \Psi^{\ast}_{n-1}(t,z),~~k=0,1,2,...\}
$$ which are orthogonal to $W_n$ by the
residue pairing
\be
\oint_{z=\iy} f(z) g(z) \frac{dz}{2\pi iz}.
\ee
Note that the plane $z^{-n}W_n$ has so-called virtual genus zero,
in the terminology of \cite{SW}; in particular, this plane
contains an element of order $1+O(z^{-1})$. The following
statement is mainly contained in \cite{AvM5}:

\bigbreak

\proclaim Proposition 2.1. The following five statements are
equivalent
\newline\noindent(i) The 1-Toda lattice equations (2.1)
\newline\noindent(ii) $\Psi$ and $\Psi^*$, with the proper
asymptotic behaviour, given by (2.4),
satisfy the bilinear identities for all
$t,t' \in \BC^{\iy }$
\be
\oint_{z=\iy}\Psi_n(t,z) \Psi^*_m(t',z) \frac{dz}{2 \pi iz}=0,
~~~\mbox{ for all }~~n>m;
\label{7}\ee
\newline\noindent(iii)
the $\tau$-vector satisfies the following bilinear identities
for all $n>m$ and $t,t' \in \BC^{\iy }$:
\begin{equation}
\oint_{z=\iy}\tau_n(t-[z^{-1}])\tau_{m+1}(t'+[z^{-1}])
e^{\sum_1^{\iy}(t_i-t'_i)z^i}
z^{n-m-1}dz=0;
\label{8}
\end{equation}
\newline\noindent(iv) The components $\tau_n$ of a $\tau$-vector
correspond to a flag of planes in $Gr$,
\begin{equation}
\supset W_{n-1}\supset
W_{n}\supset W_{n+1}\supset....
\end{equation}
\newline\noindent(v) A sequence of KP-$\tau$-functions $\tau_n$
satisfying the equations
\be\{ \tau_n
(t-[z^{-1}]),\tau_{n+1} (t)\} + z (\tau_n (t-[z^{-1}]) \tau_{n+1}
(t) - \tau_{n+1} (t-[z^{-1}])\tau_n (t)) = 0 \ee

Proof: The proof that (i) is equivalent to (ii) follows from the
methods in \cite{ASV,vM}. That (ii) is equivalent to (iii)
follows from the representation (2.4) of wave functions in
terms of
$\tau$-functions. Finally, we sketch the proof that (ii) is
equivalent to (iv). The inclusion in (iv) implies that $W_n$,
given by (2.6), is also given by
$$
W_n=\mbox{span}_{\BC}~\{\Psi_n(t,z), \Psi_{n+1}(t,z),...\};
$$
 Since each $\tau_n$ is a
$\tau$-function, we have that
$$
\oint_{z=\iy}\Psi_n(t,z) \Psi^*_{n-1}(t',z) \frac{dz}{2 \pi
iz}=0,
$$
implying that, for each $n \in \BZ$, $\Psi^*_{n-1}(t,z) \in
W_n^*$. Moreover the inclusions $...\supset
W_{n}\supset W_{n+1}\supset...$ imply, by orthogonality, the
inclusions $...\subset
W^*_{n}\subset W^*_{n+1}\subset...$, and thus
$$
W_n^*=\{ \Psi^*_{n-1}(t,z), \Psi^*_{n-2}(t,z),...\}.
$$
Since
$$
W_n\subset W_m=\left( W_m^* \right)^*,~~\mbox{ all } n\geq m,
$$
we have the orthogonality $W_n  \bot W_m^*$ by the residue
pairing (2.7) for all $n\geq m$, i.e.,
$$\oint_{z=\iy}\Psi_n(t,z) \Psi^*_{m-1}(t',z) \frac{dz}{2 \pi
iz}=0,
\mbox{ all } n\geq m.
$$
Note (ii) implies $W^{\ast}_m\subset W^{\ast}_n,~~n>m$,
hence $W_n\subset W_m,~~n>m$, yielding (iv). That (iv)$\DF
$(v) follows from proposition 1.1, by setting
$\tau_1:=\tau_n$ and $\tau_2=\tau_{n+1}$. Then (v) is equivalent
to
the inclusion property
$$
z(z^{-n-1}W_{n+1})\subset(z^{-n}W_{n}),\quad\mbox{i.e.}\quad
W_{n+1}\subset W_{n},
$$
thus ending the proof of proposition 2.1.

\qed

\bigbreak

For $m=n-2,~ t\mapsto t+[\al],~t'\mapsto t-[\al]$, the
bilinear identity (2.8) yields
\bea
0&=&
\frac{\tau_n(t+[\al])\tau_{n-1}(t-[\al])}{\tau_n(t)\tau_{n-1}(t)}
\oint_{z=\iy}\Psi_n(t+[\al],z) \Psi^*_{n-2}(t-[\al],z) \frac{dz}{
z} \nonumber  \\
&=&\frac{1}{\tau_n\tau_{n-1}}\sum_{j\geq
0}\al^j\left(p_{j+2}(\tilde \pl)-\frac{\pl}{\pl t_{j+2}}
\right)\tau_n \circ \tau_{n-1}\nonumber\\
&=&\sum_{j\geq
0}\al^j\left( (L^{j+2})_{n-1,n-1}- \frac{\pl}{\pl
t_{j+2}}\log\frac{\tau_n}{\tau_{n-1}}\right),  \nonumber
\eea
from which the following useful formula follows:
\be
 \frac{\pl }{\pl t_k}\log \frac{\tau_{n+1}}{\tau_n}=\left(
L^k  \right)_{nn}.
\ee

\bigbreak

\bigbreak

\section{Proof of Theorems 0.1 and 0.2}

\setcounter{equation}{0}

At first, we exhibit particular solutions to equation (2.11),
explained in \cite{AvM1}.

\proclaim Lemma 3.1. Particular solutions to equation
$$\{ \tau_1
(t-[z^{-1}]),\tau_2 (t)\} + z (\tau_1 (t-[z^{-1}]) \tau_2 (t)
- \tau_2 (t-[z^{-1}])\tau_1 (t)) = 0 $$
are given, for arbitrary $\lambda \in
\BC^\ast$, by pairs $(\tau_1,\tau_2)$, defined by:
\be
\tau_2 (t) = X (t,\lambda) \tau_1 (t) = e^{\sum t_i \lambda^i}
\tau_1 (t-[\lambda^{-1}]),
\ee
or
\be
\tau_1 (t) = X (-t,\lambda) \tau_2 (t) = e^{-\sum t_i \lambda^i}
\tau_2 (t+[\lambda^{-1}]).
\ee

Proof: Using
$$e^{-\sum^\iy_1 {1 \over i}({\lambda \over z})^i } = 1 -
{\lambda \over z},$$
it suffices to check that $\tau_2 (t)$
satisfies the above equation (2.11)
\bea
&&e^{-\sum t_i \lambda^i} \left(\{\tau_1
(t-[z^{-1}]),\tau_2(t)\}
 + z (\tau_1 (t-[z^{-1}])\tau_2(t) - \tau_2(t-[z^{-1}]) \tau_1
(t))\right) \nonumber\\
&&~~~~= e^{-\sum t_i \lambda^i}\{\tau_1 (t-[z^{-1}]),
e^{\sum t_i \lambda^i} \tau_1 (t-[\lambda^{-1}])\} \nonumber
\\  &&~~~~~~~~~+ z
(\tau_1 (t-[z^{-1}]) \tau_1 (t-[\lambda^{-1}]) - (1-{\lambda
\over z}) \tau_1 (t) \tau_1  (t-[z^{-1}] -
\lambda^{-1}]))\nonumber  \\ &&~~~~= \{\tau_1
(t-[z^{-1}]),\tau_1 (t-[\lambda^{-1}])\}\nonumber
\\ &&~~~~~~~~~+
(z-\lambda) (\tau_1 (t-[z^{-1}])\tau_1 (t-[\lambda^{-1}]) -
\tau_1 (t) \tau_1 (t-[z^{-1}]-[\lambda^{-1}]))\nonumber    \\
&&~~~~= 0,\nonumber
\eea
using the differential Fay identity (1.1) for the  $\tau$-function
$\tau_1$; a similar proof works for the second solution, given by
(3.2).
\qed

Proof of Theorems 0.1 and 0.2: From an arbitrary KdV
$\tau$-function, construct, for $\lb,~c,~\nu\in\BC^{\iy}$, the
following sequence of $\tau$-functions, for $n\geq 0$, as
announced in Theorem 0.1:
$$\tau_0(t)=e^{\nu_n\sum t_{iN}}\tau(c+t)
$$
\bea
\tau_n &=& e^{\nu_n\sum t_{iN}}X (t,\lambda_n) ... X
(t,\lambda_1)\tau(c+t)\nonumber\\
  &=&
{\Delta
(\lambda_1,...,\lambda_n) \over \prod^n_1
\lambda^{i-1}_i}e^{\nu_n\sum_1^{\iy} t_{iN}}
\prod^n_{k=1} e^{
\sum^{\iy}_{i=1}t_i \lambda^i_k}
 \tau (c+t-\sum^n_1 [\lambda^{-1}_i]),\nonumber
\eea
\bea
\tau_{-n} &=& e^{\nu_{-n}\sum t_{iN}}X (-t,\lambda_{-n+1}) ... X
(-t,\lambda_{0})\tau(c+t)\nonumber\\
  &=&
{\Delta
(\lambda_{0},...,\lambda_{-n+1}) \over \prod^n_1
\lambda^{i-1}_{-i+1}}e^{\nu_{-n}\sum_1^{\iy} t_{iN}}
\prod^n_{k=1} e^{-
\sum^{\iy}_{i=1}t_i \lambda^i_{-k+1}}
 \tau (c+t+\sum^n_1 [\lambda^{-1}_{-i+1}])\nonumber\\
\eea
and so, each $\tau_n$ is defined inductively by
$$
\tau_{n+1} = e^{(\nu_{n+1}-\nu_n)\sum_1^{\iy}t_{iN}}X
(t,\lambda_{n+1})
\tau_n;
$$
thus by Lemma 3.1, the functions $\tau_{n+1}$ and
$\tau_n$ are a solution of equation (v) of proposition
2.1. Therefore, by the same proposition 2.1, the
$\tau_n$'s form a $\tau$-vector of the 1-Toda
lattice. Since each $\tau_n$, except for  the exponential
factor $\exp(\nu_n\sum_1^{\iy} t_{iN})$, has the property that
$\pl\tau_n/\pl t_{iN}=0$ for
$i=1,2,...$, we have that
$$
z^N W_n\subset W_n;
$$
in particular, the representation
$$
W_n=\span\{\Psi_n(t,z),\Psi_{n+1}(t,z),...\},
$$
which follows from the inclusion $...\supset W_n\supset
W_{n+1}\supset ...$, implies that, since $L\Psi=z\Psi,$
$$
z^N\Psi_k=\sum_{j\geq k}a_j\Psi_j=(L^N\Psi)_k,
$$
and thus $L^N$ is upper-triangular. Multiplying $\tau_n$ with the
exponential factor $\exp(\nu_n\sum_1^{\iy} t_{iN})$, does not
modify the wave function $\Psi_n$, except for a factor, which is a
function of $z^N$ only.

Therefore, we conclude that the matrix $L$, defined by (2.5),
from the sequence of $\tau$-functions (3.3),
\begin{eqnarray*}
L&=&\tilde\Lb+\left(\frac{\pl}{\pl
t_1}\log\frac{\tau_{n+1}}{\tau_n}\right)_{n\in\BZ}+\left(\left(
\frac{\pl}{\pl
t_1}\right)^2\log\tau_n\right)_{n\in\BZ}\tilde\Lb^{-1}+...\\
&=&\tilde\Lb+(\lb_{n+1}+b_{n+1})_{n\in\BZ}\tilde\Lb^0+
(a_n)_{n\in\BZ}
\tilde\Lb^{-1}+...,
\end{eqnarray*}
satisfies the $1$-Toda lattice equations, where
\bea
b_{n+1}&=&\frac{\pl}{\pl
t_1}\log\frac{\tau(c+t-\sum_1^{n+1}[\lb_i^{-1}])}
{\tau(c+t-\sum_1^n[\lb_i^{-1}])}~~~\mbox{for}~~n\geq
1\nonumber\\ &=&
\frac{\pl}{\pl
t_1}\log\frac{\tau(c+t-[\lb_1^{-1}])}
{\tau(c+t)},~~~\mbox{for}~~n=0,\nonumber\\
&=&
\frac{\pl}{\pl
t_1}\log\frac{\tau(c+t+\sum_0^{n+2}[\lb_i^{-1}](1-\dt_{-1,n}))}
{\tau(c+t+\sum_0^{n+1}[\lb_i^{-1}])},~~~\mbox{for}~~n\leq -1.
\eea
\bea
a_n&=&\frac{\pl^2}{\pl
t_1^2}\log\tau(c+t-\sum_1^n[\lb_i^{-1}])~~~\mbox{for}~~n\geq
1\nonumber\\ &=&
\frac{\pl^2}{\pl
t_1^2}\log\tau(c+t)~~~\mbox{for}~~n=0\nonumber\\
&=&
\frac{\pl^2}{\pl
t_1^2}\log\tau(c+t+\sum_0^{n+1}[\lb_i^{-1}])
~~~\mbox{for}~~n\leq -1,
\eea
confirming (0.7).
Using the fact that,
in view of (2.12), the diagonal terms of $L^N$ are given by
$$
\frac{\pl}{\pl t_N}\log\frac{\tau_{n+1}}{\tau_n}=\nu_{n+1}-\nu_n,
$$
and the fact that
$$
\widetilde{\tilde\Lb}^n=(\tilde\Lb+\lb)^n=\tilde\Lb^n+
\left(\sum_1^n\lb_i\right)
\tilde\Lb^{n-1}+\left(\sum_{1\leq i\leq j\leq
n-1}\lb_i\lb_j\right)\tilde\Lb^{n-2}+...,
$$
one finds that the upper-triangular matrix $L^N$ has the
following expression:
\bea
L^N&=&\tilde\Lambda^N+\sum_1^N(\lb_j+b_j)\tilde\Lb^{p-1}+
\Biggl(\sum_0^{N-1}a_j+
\sum_{1\leq i\leq j\leq
N-1}(\lb_i+b_i)(\lb_j+b_j)\Biggr)\tilde\Lb^{N-2}\nonumber\\
& &\hspace{7cm}+...+(\nu_{n+1}
-\nu_n)_{n\in\BZ}\tilde\Lb^0\nonumber\\ & & \nonumber\\
&=&\widetilde{\tilde\Lb}^N+\left(\sum_1^N b_j\right)
\widetilde{\tilde\Lb}^{N-1}+\left(\sum_0^{N-1} a_j-\sum_1^{N-1}
(b_N-b_i)\lb_i+\sum_{1\leq i\leq j\leq
N-1}b_ib_j\right)\widetilde{\tilde\Lb}^{N-2}+...\nonumber\\
& &
\eea
in terms of $b_k$ and $a_k$ defined in (0.7), thus proving
Theorem 0.1.  \qed

To prove Theorem 0.2, note at first:
\begin{eqnarray*}
\frac{z^{n-m-1}}{\prod^n_{k=m+2}(-\lb_k)}\prod_{k=m+2}^{n}
e^{-\sum_{i=1}^{\iy}\frac{1}{i}\left(\frac{\lb_k}{z}\right)^i}
&=&\frac{z^{n-m-1}}{\prod^n_{k=m+2}(-\lb_k)}\prod_{k=m+2}^{n}
\left( 1-\frac{\lb_k}{z}  \right)\\
&=&\prod_{k=m+2}^{n}\left( 1-\frac{z}{\lb_k}  \right)\\
&=&\prod_{k=m+2}^{n}e^{-\sum_{i=1}^{\iy}
\frac{1}{i}\left(\frac{z}{\lb_k}\right)^i}\\
&=&\frac{e_q^{xzq^n}}{e_q^{xzq^{m+1}}}=
D^n e_q^{xz}~D^{m+1}\left(e_q^{xz}\right)^{-1}.
\end{eqnarray*}
The function $\tau_n$, defined in Theorem 0.1, satisfies the
bilinear identity of Theorem 0.1; therefore, using (3.3) and the
above in the computation of $\tau_n(t-[z^{-1}])$, the following
relations hold, up to a multiplicative factor depending on
$\lb$ and $\nu$:
\begin{eqnarray*}
&&\alpha(\lb,\nu)
\oint_{z=\iy}\tau_n(t-[z^{-1}])\tau_{m+1}(t'+[z^{-1}])
e^{\sum_1^{\iy}(t_i-t'_i)z^i} z^{n-m}\frac{dz}{z}\\ &&
=
\oint_{z=\iy}\tau(c(x)+t-[z^{-1}]-\sum^n_1[\lb_i^{-1}])
\tau(c(x)+t'+[z^{-1}]+\sum^{m+1}_1[\lb_i^{-1}])  \\
&&
\hspace{5cm} \prod_{k=m+2}^{n}
\left( 1-\frac{z}{\lb_k}  \right)
e^{\sum_1^{\iy}(t_i-t'_i)z^i} dz\\
&&=
\oint_{z=\iy}D^n\left( X_q(x,t,z)\tau(c(x)+t)\right)~D^{m+1}
\left(
\tilde X_q(x,t',z)\tau(c(x)+t')\right)dz=0.
\end{eqnarray*}
When $q \longrightarrow 1$, the second
expression above tends to the standard KP-bilinear
equation, upon using (0.10).  Moreover, one checks by induction,
using the first three terms in the expression for
$L$ and (2.12), that $(L^N)_+$ for $N=1,2,3,...$ has the
$q$-form (0.3).
Also, note that $a_k$ and $b_k$ can be
expressed in terms of the $D$-operator, using (0.7); to wit:
$$
b_k=\frac{\pl}{\pl
t_1}\log\frac{\tau(D^k
c+t)}{\tau(D^{k-1}c+t)},~~~~
a_k=\left(\frac{\pl}{\pl
t_1}\right)^2\log\tau\left(D^k c+t\right).
$$
So, the expression for $Q_q^N$ in Theorem 0.2 follows at once
from (3.4). The fact that
$$
-\lb_1\frac{\pl}{\pl t_1}\log
\frac{\tau(D^{j+1}c+t)}{\tau(D^jc+t)}\lrg\frac{\pl^2}{\pl
x^2}\log\tau(\bar x+t)
$$
implies that all terms in (0.10) vanish in the limit $q\lrg 1$,
except for the term $\sum_{i=0}^{N-1}\frac{\pl^2}{\pl
t_1^2}\log\tau(D^i c+t)$; so we have that
$$
\lim_{q\rg 1}Q_p^N=\left(\frac{\pl}{\pl
x}\right)^N+N\frac{\pl^2}{\pl x^2}\log\tau(\bar x+t)
\left(\frac{\pl}{\pl x}\right)^{N-2}+...,
$$
thus ending the proof of theorem 0.2.\qed

\section{Examples and vertex operators}

\setcounter{equation}{0}

The isomorphism (0.3) enables one to translate every 1-Toda
statement, having the form (0.3) into a $D$ or $D_q$ statement.
Also every $\tau$-function of the KdV hierarchy leads
automatically to a solution of $q$-KdV. For instance, by
replacing $t\mapsto c(x)+t$ in the Schur polynomials, one finds
$q$-Schur polynomials. The latter were obtained by Haine and
Iliev \cite{HI} by using the $q$-Darboux transforms; the latter
had been studied by Horozov and coworkers in \cite{BHY1,BHY2}.

The $n$-soliton solution to the KdV (for $N=2$) (for this
formulation, see \cite{ASV}),
$$
\tau (t)=\det \left(\delta_{i,j}-\frac{a_j}{y_i+y_j}
e^{-\sum_{k:\mbox{{\footnotesize odd}}} t_k (y_i^k+y_j^k)}
\right)_{1\leq i, j\leq n},
$$
leads to a $q$-soliton by the shift $t \mapsto c(x)+t$, with
$c(x)$ as in (0.8), namely
$$
\tau(x,t)=\det \left(\delta_{ij}-a_j \frac{\left(e_q^{xy_i}
e_q^{xy_i}\right)^{-1}}{y_i+y_j} e^{-\sum_{k=1}^{\iy}
t_k(y_i^k+y_j^k)}\right)_{1\leq i,j\leq n}.
$$
Moreover the vertex operator for the 1-Toda lattice is a
reduction of the 2-Toda lattice vertex operator (see
\cite{AvM4}), given by
\begin{eqnarray*}
\BX(t,y,z)&=&-\chi^{\ast}(z) X(-t,z) X(t,y) \chi(y)\\
&=& \frac{z}{y-z}
e^{\sum_1^{\iy} t_i (y^i-z^i)} e^{-\sum_1^{\iy}(y^{-i}-z^{-i})
\frac{1}{i}
\frac{\pl}{\pl t_i}}\left(\frac{y^n}{z^n}\right)_{n\in\BZ};
\end{eqnarray*}
in particular, if $\tau$ is a 1-Toda vector, then
$ a\tau+b\BX(t,y,z)\tau$ is a 1-Toda vector as well.  Using the
dictionary, this leads to
$q$-vertex operators
$$
\BX_q (x,t;y,z)=e_q^{xy}(e_q^{xz})^{-1} e^{\sum t_i (y^i-z^i)}
e^{-\sum(y^{-i}-z^{-i}) \frac{1}{i}
\frac{\pl}{\pl t_i}}
\quad \mbox{for $q$-KP},
$$
and, for any $N$th root $\omega$ of 1,
$$
\BX_q (x,t;z)= e_q^{x\om z} (e_q^{xz})^{-1} e^{\sum t_i z^i
(\om^i-1)} e^{-\sum z^{-i}(\om^{-i}-1) \frac{1}{i}\frac{\pl}{\pl
t_i}}
\quad \mbox{for $q$-KdV,}
$$
having the typical vertex operator properties.



\begin{thebibliography}{99}

\bibitem{AvM1} M.~ Adler and P.~ van Moerbeke: {\em Birkhoff
strata, B\"acklund transformations and limits of isospectral
operators }, Adv. in  Math., {\bf 108} 140--204 (1994).



\bibitem{AvM4} M. Adler and P. van Moerbeke: {\em The spectrum of
coupled random matrices}, preprint (1997).

\bibitem{AvM5} M. Adler and P. van Moerbeke: {\em The full one-Toda
lattice}, preprint (1997).


\bibitem{ASV} M.~Adler, T.~Shiota and P.~ van Moerbeke : {\em
A Lax representation for the vertex operator and the central
extension}, Comm. Math. Phys. {\bf 171}, 547--588, (1997).


\bibitem{BHY1} B.~ Bakalov, E.~ Horozov and M.~ Yakimov: {\em
Bispectral algebras of commuting ordinary differential operators },
Comm. Math. Phys., to appear.

\bibitem{BHY2} B.~ Bakalov, E.~ Horozov and M.~ Yakimov: {\em
General methods for constructing bispectral operators }, Phys.
Letters A, {\bf 222} 59--66 (1996).

\bibitem{F} E.~ Frenkel: {\em Deformations of the KdV hierarchy
and related soliton equations }, Int. Math. Res. Notices, {\bf 2}
55--76 (1996).


\bibitem{G} D.~ Gieseker: {\em The Toda hierarchy and the KdV
hierarchy }, preprint, alg-geom/9509006.


\bibitem{HI} L.~ Haine and P. ~Iliev: {\em The bispectral
property of a $q$-deformation of the Schur polynomials and the
$q$-KdV hierarchy }, J. of Phys. A; math.and gen,
(1997), to appear.

\bibitem{KLR} B.~ Khesin, V.~ Lyubashenko and C.~Roger: {\em
Extensions and contractions of the Lie algebra of
$q$-Peudodifferential symbols on the circle }, J. of functional
analysis, {\bf 143} 55--97 (1997).

\bibitem{K} B. A.~ Kupershmidt: {\em Discrete Lax equations and
differential-difference calculus }, Ast\'erisque, {\bf 123}
(1985).


\bibitem{SW} G. ~Segal, G. ~Wilson: {\em Loop groups and equations
of KdV type}, Publ. Math. IHES {\bf 61}, 5--65 (1985).

\bibitem{T} K.~Takasaki, T. ~Takebe: {\em Integrable
hierarchies and dispersionless limit}, Reviews in Math. Phys.
{\bf 7},743--808  (1995) .



\bibitem{vM} P.~van Moerbeke: Integrable foundations of string
theory, in Lecures on Integrable systems,  Proceedings of the
CIMPA-school, 1991, Ed.: O. Babelon, P. Cartier, Y.
Kosmann-Schwarzbach, World scientific, pp 163--267 (1994).


\end{thebibliography}
\end{document}